\documentclass[twocolumn,english]{revtex4}
\usepackage[T1]{fontenc}
\usepackage[latin1]{inputenc}
\usepackage{amsmath}
\usepackage{amssymb}
\usepackage{graphicx}
\usepackage{esint}

\makeatletter
\@ifundefined{textcolor}{}
{%
 \definecolor{BLACK}{gray}{0}
 \definecolor{WHITE}{gray}{1}
 \definecolor{RED}{rgb}{1,0,0}
 \definecolor{GREEN}{rgb}{0,1,0}
 \definecolor{BLUE}{rgb}{0,0,1}
 \definecolor{CYAN}{cmyk}{1,0,0,0}
 \definecolor{MAGENTA}{cmyk}{0,1,0,0}
 \definecolor{YELLOW}{cmyk}{0,0,1,0}
}

\usepackage{amsfonts}\usepackage{babel}

\usepackage{babel}

\makeatother

\usepackage{babel}
\begin{document}

\title{Chirality asymptotic behavior and non-Markovianity in quantum walks
on a line}

\author{Margarida Hinarejos$^{1}$, Carlo Di Franco$^{2}$, Alejandro Romanelli$^{3}$,
and Armando Pérez$^{1}$ }

\affiliation{$^{1}$Departament de F\'{i}sica Teòrica and IFIC, Universitat de
València-CSIC \\
 Dr. Moliner 50, 46100-Burjassot, Spain\\
 $^{2}$ Centre for Theoretical Atomic, Molecular and Optical Physics,
\\
 School of Mathematics and Physics, Queen's University, Belfast,\\
 BT7 1NN, United Kingdom\\
 $^{3}$ Instituto de F\'{i}sica, Facultad de Ingenier\'{i}a, Universidad
de la República, \\
 C.C. 30, C.P. 11000, Montevideo, Uruguay}
\begin{abstract}
We investigate the time evolution of the chirality reduced density
matrix for a discrete-time quantum walk on a one-dimensional lattice,
which is obtained by tracing out the spatial degree of freedom. We
analyze the standard case, without decoherence, and the situation
where decoherence appears in the form of broken links in the lattice.
By examining the trace distance for possible pairs of initial states
as a function of time, we conclude that the evolution of the reduced
density matrix is non-Markovian, in the sense defined in {[}H. P.
Breuer, E. M. Laine, and J. Piilo, Phys. Rev. Lett. \textbf{103},
210401 (2009){]}. As the level of noise increases, the dynamics approaches
a Markovian process. The highest non-Markovianity corresponds to the
case without decoherence. The reduced density matrix tends always
to a well-defined limit that we calculate, but only in the decoherence-free
case this limit is non-trivial. 
\end{abstract}
\maketitle

\section{Introduction}

Markov approximation is a valuable and powerful tool for studying
the dynamics of an open system interacting with its environment. It
holds when full predictions on the future evolution of the system
can be obtained by only knowing its present state, and no further
knowledge of its past is required. The classical random walk is an
example of Markovian process that has found applications in many fields.
In quantum mechanics, important physical processes leading to decoherence
can be analyzed by means of simple Markovian models. For instance,
in quantum optics, the time evolution of an open system characterized
by a non-unitary behavior can be described by a master equation written
generally in the form of a Lindblad equation \cite{BreuerPetruccione}.

In quantum information theory, the discrete-time quantum walk (QW)
on a line has been studied as a natural generalization of the classical
random walk \cite{Aharonov}. In this context, it has been shown in
detail how the unitary quantum mechanical evolution of the QW can
be separated into Markovian and interference terms \cite{alejo2003,alejo2004}.
The Markovian terms responsible for the diffusion obey a master equation,
while the others include the interference terms needed to preserve
the unitary character of the evolution. This approach provides an
intuitive framework which becomes useful for analyzing the behavior
of quantum systems in which decoherence plays a central role. In other
words, this formalism shows in a transparent form that the primary
effect of decoherence here is to make the interference terms negligible
in the evolution equation, and then the Markovian behavior is immediately
obtained.

In this scenario, it is important to find a way to evaluate how non-Markovian
a quantum system is. Ref. \cite{Breuer2009} has proposed a general
measure for the degree of non-Markovian behavior in open quantum systems.
This measure is based on the trace distance, which quantifies the
distinguishability of quantum states, and can be interpreted in terms
of the information flow between the open system and its environment.
The measure takes nonzero values whenever there is a flow of information
from the environment back to the open system, and it has already been
used in different contexts \cite{BreuerUses}.

On the other hand, the asymptotic behavior of the QW has been recently
investigated focusing on the chirality reduced density matrix, obtained
when the position degree of freedom is traced out \cite{alejo2010,alejo2012,alejo2013a,alejo2013b}.
This matrix has a long-time limit that depends on the initial conditions.
One finds thus the following situation: the dynamical evolution of
the QW is a unitary process, however the asymptotic behavior of the
reduced density matrix has some properties which are characteristic
of a diffusive Markov process. This asymptotic behavior allows to
amalgamate concepts such as thermodynamic equilibrium with the idea
of a system that follows a unitary evolution. Refs. \cite{alejo2012,alejo2013b}
have developed a thermodynamic theory to describe the QW equilibrium
between the position and chirality degrees of freedom. They have shown
that it is possible to introduce the concept of temperature for an
isolated quantum system that evolves in a composite Hilbert space
(i.e. the tensor product of several subspaces). Additionally, Ref.
\cite{alejo2012} has shown that the transient behavior towards thermodynamic
equilibrium is described by a master equation with a time-dependent
population rate. 

In this paper we study the asymptotic QW behavior with and without
decoherence and exploit the measure proposed in Ref. \cite{Breuer2009}
to evaluate its non-Markovianity. We show that, without decoherence,
the reduced density matrix dynamics has a clear time dependence that
gives rise to a non-Markovian behavior, as we later confirm by examining
the short-time evolution of the trace distance between pairs of states.
The chirality density matrix has a well-defined limit that can be
calculated in terms of the initial conditions. This corresponds, when
comparing the evolution of two different initial states, to a reduced
asymptotic trace distance. The introduction of decoherence translates,
as the long-term limit is concerned, into a trivial result, since
all states evolve towards the maximally decohered state (proportional
to the identity matrix).

The evolution during the first time steps of the QW features an interesting
phenomenon, i.e. the presence of oscillations in the trace distance
between pairs of states, which is interpreted as a signature of a
non-Markovian process. Such oscillations occur both with and without
decoherence, even though they become more and more attenuated as the
level of noise increases. In agreement with our observations for the
asymptotic limit, the trace distance tends to zero when decoherence
affects the system.

This paper is organized as follows. In Sect. II we introduce the basic
features of the QW and obtain the asymptotic limit for the reduced
density matrix in the chiral space. In Sect. III we recast the QW
in the form of a map equation for the Generalized Chiral Distribution
(GCD), i.e. the diagonal terms of the reduced density matrix, in connection
with the non-Markovianity of the time evolution for the reduced system.
The asymptotic limit under the effect of decoherence is addressed
in Sect. IV. In Sect. V we discuss the short-time behavior, where
non-Markovian effects clearly manifest as oscillations of the trace
distance between pairs of states. Sect. VI summarizes our main results.

\section{Asymptotic reduced density matrix for the QW}

The standard QW corresponds to the discrete (both in time and in space)
evolution of a one-dimensional quantum system (the walker) in a direction
which depends on an additional degree of freedom, the chirality, with
two possible states: ``left\textquotedblright{} $|L\rangle$ or ``right\textquotedblright{}
$|R\rangle$. The global Hilbert space of the system is the tensor
product $H_{s}\otimes H_{c}$. $H_{s}$ is the Hilbert space associated
to the motion on the line, and it is spanned by the basis $\{|x\rangle:x\in\mathbb{Z}\}$.
$H_{c}$ is the chirality (or coin) Hilbert space, defined as a two-dimensional
space that can correspond, for example, to a spin 1/2 particle, or
to a 2-level energy system. Let us call $T_{-}$ ($T_{+}$) the operators
in $H_{s}$ that move the walker one site to the left (right), and
$|L\rangle\langle L|$ and $|R\rangle\langle R|$ the chirality projector
operators in $H_{c}$. We consider the unitary transformation 
\begin{equation}
U(\theta)=\left\{ T_{-}\otimes|L\rangle\langle L|+T_{+}\otimes|R\rangle\langle R|\right\} \circ\left\{ I\otimes K(\theta)\right\} ,\label{Ugen}
\end{equation}
where $K(\theta)=\sigma_{z}e^{-i\theta\sigma_{y}}$, $\theta\in\left[0,\pi/2\right]$
is a parameter defining the bias of the coin toss, $I$ is the identity
operator in $H_{s}$, and $\sigma_{y}$ and $\sigma_{z}$ are Pauli
matrices acting on $H_{c}$. The effect of the unitary operator $U(\theta)$
on the state of the system in one time step $\tau$ is $|\Psi(t+\tau)\rangle=U(\theta)|\Psi(t)\rangle$.
The state vector can be expressed as the spinor 
\begin{equation}
|\Psi(t)\rangle=\sum\limits _{x=-\infty}^{\infty}\left[\begin{array}{c}
a_{x}(t)\\
b_{x}(t)
\end{array}\right]|x\rangle,\label{spinor}
\end{equation}
where the upper (lower) component is associated to the left (right)
chirality. The unitary evolution implied by Eq.(\ref{Ugen}) can be
written as the map 
\begin{align}
a_{x}(t+\tau) & =a_{x+1}(t)\,\cos\theta\,+b_{x+1}(t)\,\sin\theta,\label{mapa0}\\
b_{x}(t+\tau) & =a_{x-1}(t)\,\sin\theta\,-b_{x-1}(t)\,\cos\theta.\label{mapa}
\end{align}
In this paper we select $\theta=\frac{\pi}{4}$ to obtain an unbiased
coin (Hadamard coin).

The density matrix of the quantum system is 
\begin{equation}
\rho(t)=|\Psi(t)\rangle\langle\Psi(t)|.\label{uno}
\end{equation}
To study the QW time dependence on the initial conditions, we take
the initial state of the walker as sharply localized at the origin
with arbitrary chirality, thus 
\begin{equation}
|\Psi(0)\rangle=|0\rangle\otimes\left\vert \Phi_{0}\right\rangle ,\label{psi0}
\end{equation}
where 
\begin{equation}
\left\vert \Phi_{0}\right\rangle =\binom{\cos\frac{\gamma}{2}}{e^{-i\varphi}\sin\frac{\gamma}{2}},\label{phsi0}
\end{equation}
with $\gamma\in\left[0,\pi\right]$ and $\varphi\in\left[0,2\pi\right]$
defining a point on the unit three-dimensional Bloch sphere. In this
case the initial density matrix is 
\begin{eqnarray}
\rho(0) & = & \left\vert 0\right\rangle \left\langle 0\right\vert \otimes\left\vert \Phi_{0}\right\rangle \left\langle \Phi_{0}\right\vert ,\label{opera0}
\end{eqnarray}
where 
\begin{equation}
\left\vert \Phi_{0}\right\rangle \left\langle \Phi_{0}\right\vert =\left(\begin{array}{cc}
^{(\cos\frac{\gamma}{2})^{2}} & \frac{e^{i\varphi}}{2}\sin\gamma\\
\frac{e^{-i\varphi}}{2}\sin\gamma & ^{(\sin\frac{\gamma}{2})^{2}}
\end{array}\right).\label{opera}
\end{equation}
In order to use the affine map approach \cite{Brun,Mostafa}, Eq.(\ref{opera})
can be transformed to express the two-by-two matrix as a four-dimensional
column vector, obtaining 
\begin{eqnarray}
\left\vert \Phi_{0}\right\rangle \left\langle \Phi_{0}\right\vert  & = & r_{0}I+r_{1}\sigma_{1}+r_{2}\sigma_{2}+r_{3}\sigma_{3}\nonumber \\
 & = & \left(\begin{array}{c}
r_{0}\\
r_{1}\\
r_{2}\\
r_{3}
\end{array}\right)=\frac{1}{2}\left(\begin{array}{c}
1\\
\cos\varphi\sin\gamma\\
-\sin\varphi\sin\gamma\\
\cos\gamma
\end{array}\right),
\end{eqnarray}
where $\sigma_{i}$ with ${i=1,2,3}$ are the Pauli matrices, and
\begin{equation}
r_{i}=\frac{1}{2}\mathrm{tr}(\left\vert \Phi_{0}\right\rangle \left\langle \Phi_{0}\right\vert \sigma_{i}).\label{traza}
\end{equation}

The reduced density operator is defined as 
\begin{equation}
\rho_{c}(t)=\mathrm{tr_{s}}(\rho(t))=\sum_{x=-\infty}^{\infty}\langle x|\rho(t)|x\rangle,\label{dos}
\end{equation}
where the partial trace is taken over the positions. Following the
method introduced in Ref. \cite{Brun} and generalized in Ref. \cite{Mostafa},
Eq.(\ref{dos}) can be transformed into 
\begin{equation}
\rho_{c}(t)=\int\limits _{-\pi}^{\pi}\frac{dk}{2\pi}\mathit{L}_{k}^{t}\left\vert \Phi_{0}\right\rangle \left\langle \Phi_{0}\right\vert ,\label{tres}
\end{equation}
where $\mathit{L}_{k}$ is the superoperator defined as 
\begin{equation}
\mathit{L}_{k}=\left(\begin{array}{cccc}
1 & 0 & 0 & 0\\
0 & 0 & \sin2k & \cos2k\\
0 & 0 & -\cos2k & \sin2k\\
0 & 1 & 0 & 0
\end{array}\right).\label{Lkk}
\end{equation}
In order to obtain the eigenvalues of $\mathit{L}_{k}$, it is necessary
to find the eigenvalues of the following associated matrix 
\begin{equation}
\mathit{M}_{k}=\left(\begin{array}{ccc}
0 & \sin2k & \cos2k\\
0 & -\cos2k & \sin2k\\
1 & 0 & 0
\end{array}\right).\label{mk}
\end{equation}
The eigenvalues of Eq.(\ref{mk}) are 
\begin{equation}
\lambda_{1}=1,\text{ \ }\lambda_{2}=e^{i(\alpha+\pi)},\text{ \ }\lambda_{2}=e^{-i(\alpha+\pi)},
\end{equation}
where 
\begin{equation}
\cos\alpha=\frac{1}{2}(1+\cos2k)=(\cos k)^{2}.
\end{equation}
The corresponding eigenvectors are 
\begin{eqnarray}
\overset{\rightarrow}{v_{1}} & = & \left(\begin{array}{c}
v_{11}\\
v_{21}\\
v_{31}
\end{array}\right)\nonumber \\
 & = & \frac{\sqrt{2}\cos k}{\sqrt{3+\cos2k}}\left(\begin{array}{c}
1\\
(1-\cos2k)/\sin2k\\
1
\end{array}\right)\!,
\end{eqnarray}
\begin{eqnarray}
\overset{\rightarrow}{v_{2}} & = & \left(\begin{array}{c}
v_{12}\\
v_{22}\\
v_{32}
\end{array}\right)\nonumber \\
 & = & \frac{1}{N_{2}}\left(\begin{array}{c}
e^{i(\alpha+\pi)}\\
-(e^{i(\alpha+\pi)}-2\cos2k)/(2\sin2k)\\
1
\end{array}\right)\!,
\end{eqnarray}
\begin{eqnarray}
\overset{\rightarrow}{v_{3}} & = & \left(\begin{array}{c}
v_{13}\\
v_{23}\\
v_{33}
\end{array}\right)\nonumber \\
 & = & \!\frac{1}{N_{3}}\!\left(\begin{array}{c}
e^{-i(\alpha+\pi)}\\
\!-(e^{-i(\alpha+\pi)}\!-\!2\cos2k)/(2\sin2k)\!\\
1
\end{array}\right)\!,
\end{eqnarray}
where $N_{2}$ and $N_{3}$ are normalization factors. It is now straightforward
to obtain $\left(\mathit{L}_{k}\right)^{t}$ using the diagonal expression
for $\mathit{L}_{k}$, that is 
\begin{equation}
L_{k}=B\left(\begin{array}{cccc}
1 & 0 & 0 & 0\\
0 & 1 & 0 & 0\\
0 & 0 & e^{it(\alpha+\pi)} & 0\\
0 & 0 & 0 & e^{-it(\alpha+\pi)}
\end{array}\right)B^{\dagger}.\label{Lt}
\end{equation}
Here, $B$ is the eigenvector matrix 
\begin{equation}
B=\left(\begin{array}{cccc}
1 & 0 & 0 & 0\\
0 & v_{11} & v_{12} & v_{13}\\
0 & v_{21} & v_{22} & v_{23}\\
0 & v_{31} & v_{32} & v_{33}
\end{array}\right),\label{bb}
\end{equation}
and $B^{\dagger}$ its transposed conjugate. Substituting Eq. (\ref{bb})
into Eq. (\ref{Lt}) and exploiting the stationary phase theorem to
neglect the oscillatory terms $e^{\pm it(\alpha+\pi)}$ when time
goes to infinity, one finds the following asymptotic equation 
\begin{eqnarray}
\left(\mathit{L}_{k}\right)^{t} & \longrightarrow & \left(\begin{array}{cccc}
1 & 0 & 0 & 0\\
0 & \left\vert v_{11}\right\vert ^{2} & v_{11}v_{21}^{\ast} & v_{11}v_{31}^{\ast}\\
0 & v_{21}v_{11}^{\ast} & \left\vert v_{21}\right\vert ^{2} & v_{21}v_{31}^{\ast}\\
0 & v_{31}v_{11}^{\ast} & v_{31}v_{21}^{\ast} & \left\vert v_{31}\right\vert ^{2}
\end{array}\right).
\end{eqnarray}
The reduced density matrix in the asymptotic regime, $\overset{\sim}{\rho_{c}}$,
can be calculated using Eq.(\ref{tres}) as 
\begin{eqnarray}
\overset{\sim}{\rho_{c}}\equiv\lim_{t\rightarrow\infty}\rho_{c}(t) & =\lim_{t\rightarrow\infty} & \int\limits _{-\pi}^{\pi}\frac{dk}{2\pi}\mathit{L}_{k}^{t}\left\vert \Phi_{0}\right\rangle \left\langle \Phi_{0}\right\vert .\label{rr}
\end{eqnarray}
In order to work out the latter expression, it is necessary to solve
the following integrals 
\begin{eqnarray}
\int\limits _{-\pi}^{\pi}\frac{\left\vert v_{11}\right\vert ^{2}}{2\pi}dk & = & 1-\frac{1}{\sqrt{2}},\\
\int\limits _{-\pi}^{\pi}\frac{\left\vert v_{21}\right\vert ^{2}}{2\pi}dk & = & \sqrt{2}-1,\\
\int\limits _{-\pi}^{\pi}\frac{\left\vert v_{31}\right\vert ^{2}}{2\pi}dk & = & 1-\frac{1}{\sqrt{2}},\\
\int\limits _{-\pi}^{\pi}\frac{v_{11}v_{21}^{\ast}}{2\pi}dk & = & \int_{-\pi}^{\pi}\frac{v_{11}v_{31}^{\ast}}{2\pi}dk\nonumber \\
 & = & \int\limits _{-\pi}^{\pi}\frac{v_{21}v_{31}^{\ast}}{2\pi}dk=0.
\end{eqnarray}
Therefore, we obtain analytically the QW reduced density matrix in
the asymptotic regime, 
\begin{eqnarray}
\overset{\sim}{\rho_{c}} & = & \left(\begin{array}{c}
r_{0}\\
\left(1-\frac{1}{\sqrt{2}}\right)\left(r_{1}+r_{3}\right)\\
\left(\sqrt{2}-1\right)r_{2}\\
\left(1-\frac{1}{\sqrt{2}}\right)\left(r_{1}+r_{3}\right)
\end{array}\right)\nonumber \\
 & = & \frac{1}{2}\left(\begin{array}{c}
1\\
\left(1-\frac{1}{\sqrt{2}}\right)\left(\cos\varphi\sin\gamma+\cos\gamma\right)\\
\left(\sqrt{2}-1\right)\sin\varphi\sin\gamma\\
\left(1-\frac{1}{\sqrt{2}}\right)\left(\cos\varphi\sin\gamma+\cos\gamma\right)
\end{array}\right).
\end{eqnarray}
Returning to the $2\times2$ matrix formalism, the reduced density
matrix in the asymptotic regime can finally be written as 
\begin{equation}
\overset{\sim}{\rho_{c}}=\left(\begin{array}{cc}
\Pi_{\mathit{L}} & \mathit{Q}_{0}\\
\mathit{Q}_{0}^{\ast} & \Pi_{\mathit{R}}
\end{array}\right),
\end{equation}
where 
\begin{eqnarray}
\Pi_{\mathit{L}} & = & \frac{1}{2}\left[1+\left(1-\frac{1}{\sqrt{2}}\right)\left(\cos\varphi\sin\gamma+\cos\gamma\right)\right],\nonumber \\
\Pi_{\mathit{R}} & = & \frac{1}{2}\left[1-\left(1-\frac{1}{\sqrt{2}}\right)\left(\cos\varphi\sin\gamma+\cos\gamma\right)\right],\nonumber \\
\mathit{Q}_{0} & = & \frac{1}{2}\left(1-\frac{1}{\sqrt{2}}\right)\left[\left(\cos\varphi\sin\gamma+\cos\gamma\right)\right.\nonumber \\
 &  & \left.-i\sqrt{2}\sin\varphi\sin\gamma\right].\label{q0}
\end{eqnarray}

\section{QW map equation}

The aim of this Section is to connect the reduced density matrix of
the standard (decoherence-free) QW with its non-Markovian behavior.
Using Eq. (\ref{spinor}), Eq. (\ref{uno}), and Eq. (\ref{dos}),
the reduced density matrix is expressed as 
\begin{equation}
\rho_{c}(t)=\left(\begin{array}{cc}
P_{L}(t) & \mathit{Q}(t)\\
\mathit{Q}^{\ast}(t) & P_{R}(t)
\end{array}\right),\label{chiral_t}
\end{equation}
where 
\begin{align}
P_{L}(t) & =\sum_{k=-\infty}^{\infty}\left\vert a_{k}(t)\right\vert ^{2},\\
P_{R}(t) & =\sum_{k=-\infty}^{\infty}\left\vert b_{k}(t)\right\vert ^{2},\label{chirality}
\end{align}
\begin{equation}
Q(t)\equiv\sum_{k=-\infty}^{\infty}a_{k}(t)b_{k}^{\ast}(t).\label{qdet}
\end{equation}
The global chirality distribution (GCD) is defined as the distribution
\begin{equation}
\left[\begin{array}{c}
P_{L}(t)\\
P_{R}(t)
\end{array}\right],
\end{equation}
with $P_{R}(t)+P_{L}(t)=1$.

It is shown in Ref. \cite{alejo2010} that the GCD satisfies the following
map 
\begin{align}
\left[\begin{array}{c}
P_{L}(t+1)\\
P_{R}(t+1)
\end{array}\right] & =\left(\begin{array}{cc}
\cos^{2}\theta & \sin^{2}\theta\\
\sin^{2}\theta & \cos^{2}\theta
\end{array}\right)\left[\begin{array}{c}
P_{L}(t)\\
P_{R}(t)
\end{array}\right]\nonumber \\
 & +\mathrm{Re}\left[Q(t)\right]\sin2\theta\left[\begin{array}{c}
1\\
-1
\end{array}\right].\label{master}
\end{align}
From this equation, it is straightforward to observe that, if the
``intereference term\textquotedblright{} 
\begin{equation}
\mathrm{Re}\left[Q(t)\right]\sin2\theta\left[\begin{array}{c}
1\\
-1
\end{array}\right]\label{mam2}
\end{equation}
in Eq. (\ref{master}) can be neglected, the time evolution of the
GCD is described by a Markovian process in which the two-dimensional
matrix 
\begin{equation}
\left(\begin{array}{cc}
\cos^{2}\theta & \sin^{2}\theta\\
\sin^{2}\theta & \cos^{2}\theta
\end{array}\right)\label{mam}
\end{equation}
can be interpreted as the corresponding transition probability matrix
for a Markov chain, since it satisfies the necessary requirements:
all its elements are positive and the sum over the elements of any
column or row is equal to one.

Only if $\mathrm{Re}[Q(t)]$ vanished, the behavior of the GCD could
be described as a classical Markovian process. However $Q(t)$, together
with $P_{L}(t)$ and $P_{R}(t)$, are time-dependent functions. This
implies that the map defined by Eq.(\ref{master}) does not correspond
to a Markovian process. In Sect. V, we analyze this feature in more
detail, and quantify how much the QW departs from a Markovian process.

In spite of the time dependence manifested by Eq. (\ref{master}),
the GCD does possess a long-time limiting value, as obtained in previous
Section. Eq.(\ref{master}) can be used to derive a consistency condition
relating $\Pi_{L}$, $\Pi_{R}$, and $Q_{0}$, by taking the limit
$t\rightarrow\infty$. One then obtains 
\begin{equation}
\left[\begin{array}{c}
\Pi_{L}\\
\Pi_{R}
\end{array}\right]=\frac{1}{2}\left[\begin{array}{c}
1+2\mathrm{Re}(Q_{0})/\tan\theta\\
1-2\mathrm{Re}(Q_{0})/\tan\theta
\end{array}\right].\label{estacio}
\end{equation}
When $\theta=\pi/4$, Eq.(\ref{estacio}) agrees with the expression
given by Eq.(\ref{q0}). This interesting result for the QW shows
that the long-time probability to find the system with left or right
chirality only depends on the asymptotic interference term. Although
the dynamical evolution of the QW is unitary, the evolution of its
GCD has an asymptotic limit, a feature which is characteristic of
a diffusive behavior. This situation is even more surprising if we
compare our case with the case of the QW on finite graphs \cite{Aharonov},
where it is shown that there is no convergence to a stationary distribution.
In order to quantify how much the asymptotic limit keeps track of
the initial state, we use the trace distance 
\[
D(\rho_{1},\rho_{2})=\frac{1}{2}tr\arrowvert\rho_{1}-\rho_{2}|,
\]
which gives us a measure for the distinguishability of two quantum
states. Here, $\left\vert \rho\right\vert =\sqrt{\rho^{\dagger}\rho}$.
We calculate this quantity for two reduced density matrices (in the
chiral space) that correspond to two different initial states of Eq.(\ref{opera}).
Following the notation defined in Eq.(\ref{chiral_t}), we write 
\begin{equation}
\rho_{1}(t)-\rho_{2}(t)=\left(\begin{array}{cc}
P_{1L}(t)-P_{2L}(t) & \mathit{Q_{1}}(t)-\mathit{Q_{2}}(t)\\
\mathit{Q}_{1}^{\ast}(t)-\mathit{Q}_{2}^{\ast}(t) & P_{1R}(t)-P_{2R}(t)
\end{array}\right).\label{rho12_t}
\end{equation}

\begin{figure}
\begin{minipage}[t]{1\columnwidth}%
\includegraphics[width=4.5cm]{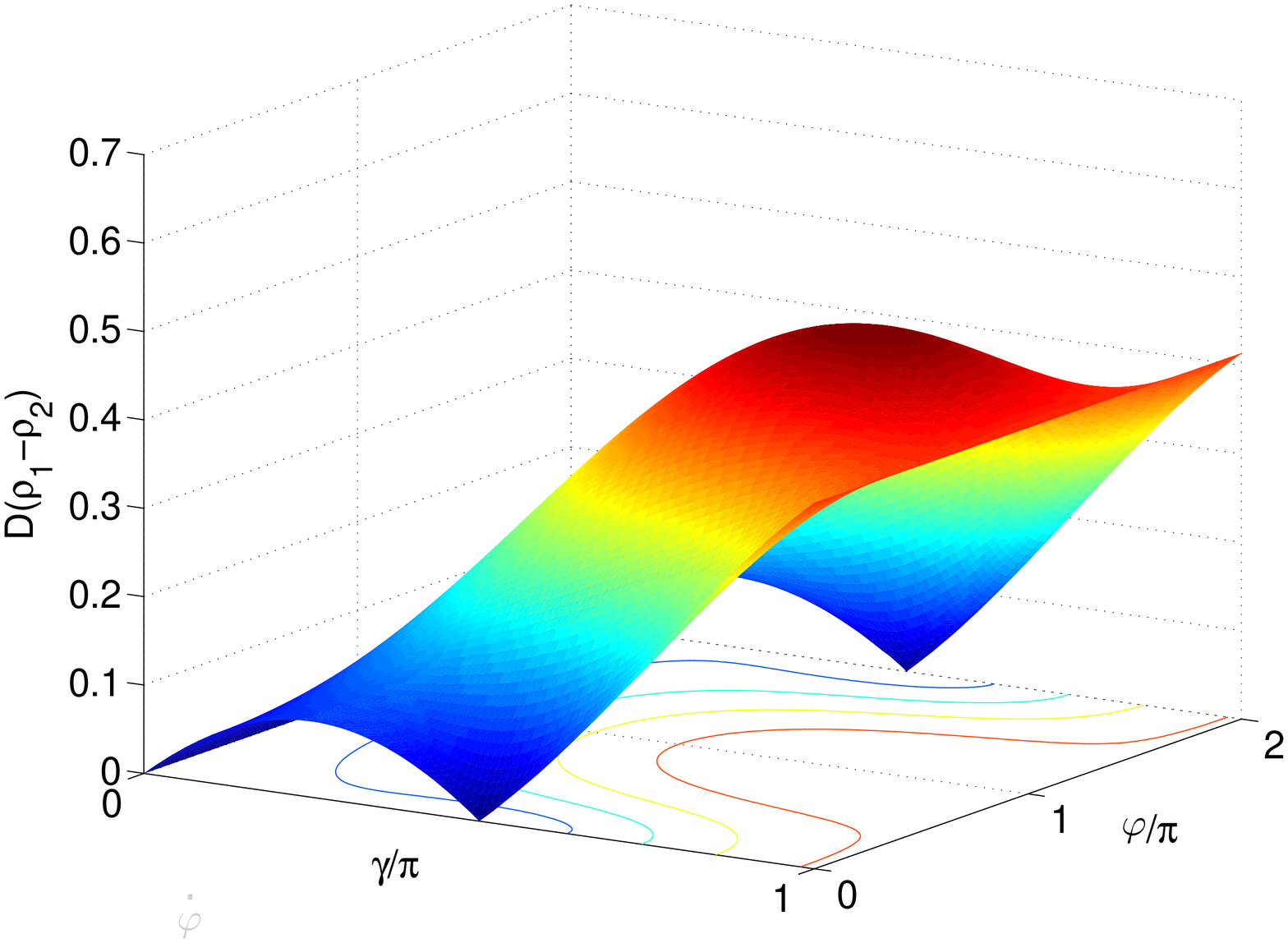}\includegraphics[width=4.5cm]{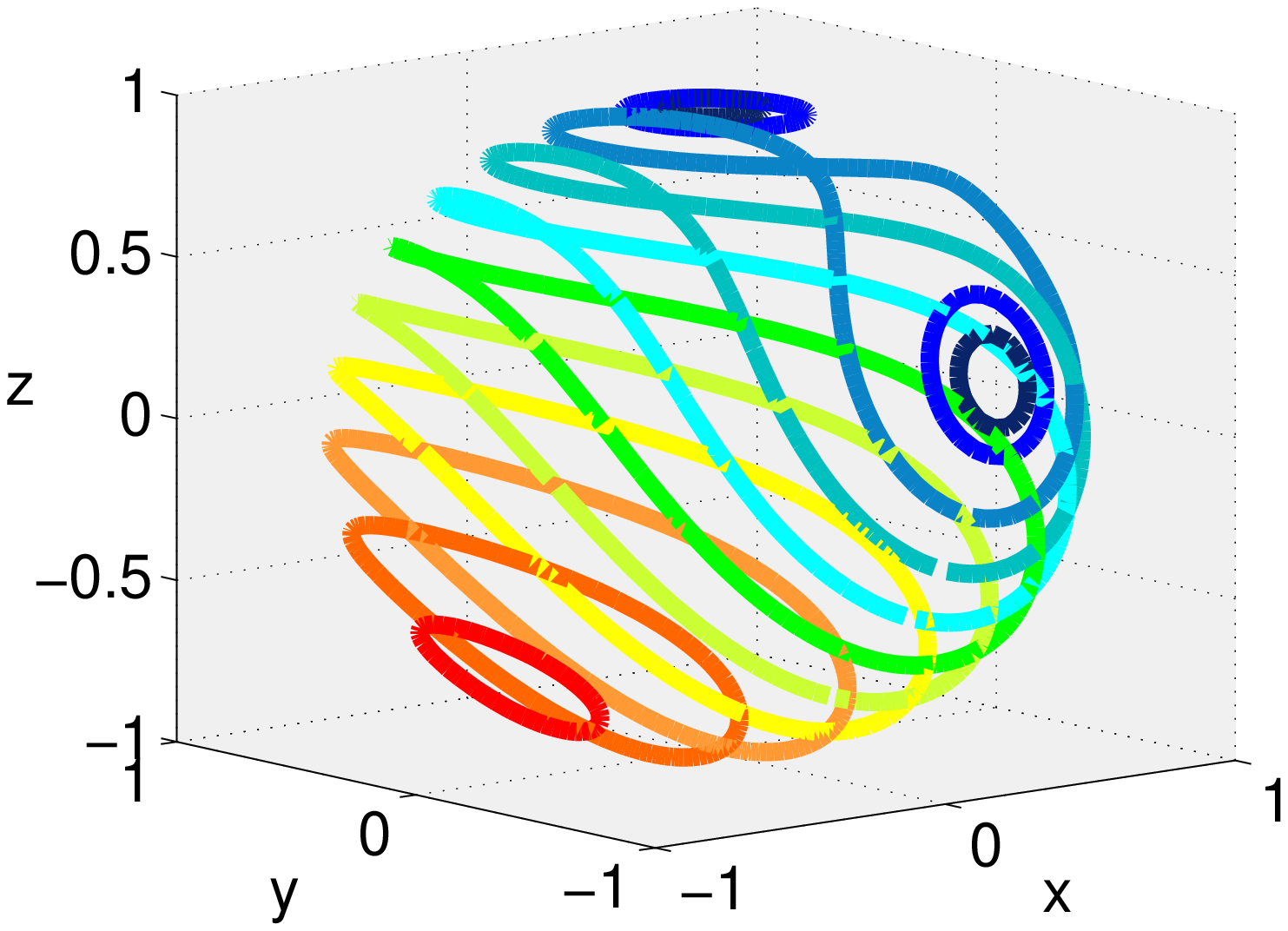}%
\end{minipage}

\caption{Left panel: Asymptotic trace distance as a function of the angles
$\gamma$ and $\varphi$, representing the initial conditions of $\rho_{2}$.
The initial conditions of $\rho_{1}$ are given by $\gamma=0$. Right
panel: The contour levels corresponding to to the left panel are mapped
to the Bloch sphere, using the same color convention. }

\label{fa} 
\end{figure}

\begin{figure}
\begin{minipage}[t]{1\columnwidth}%
\includegraphics[width=4.5cm]{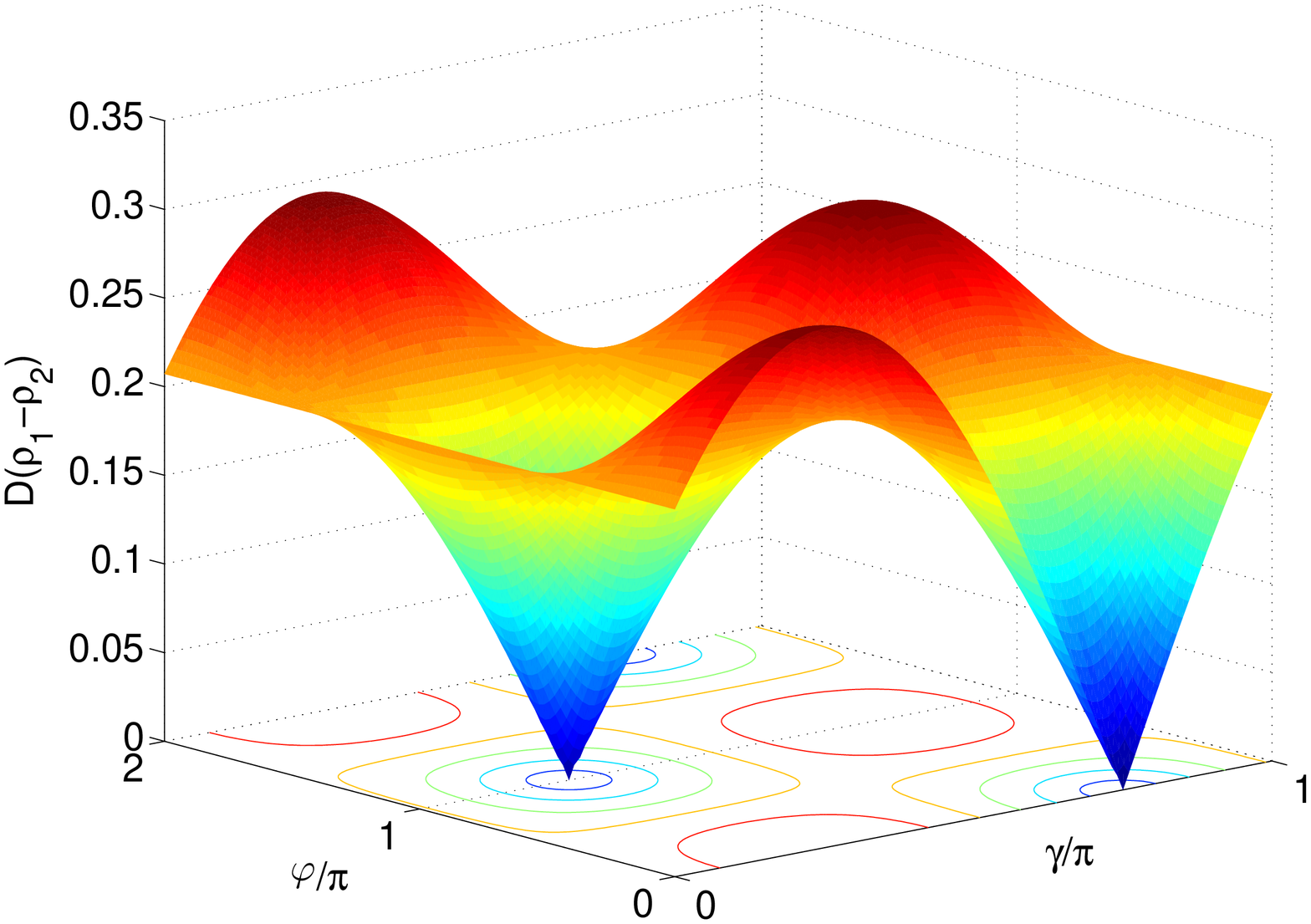}\includegraphics[width=4.5cm]{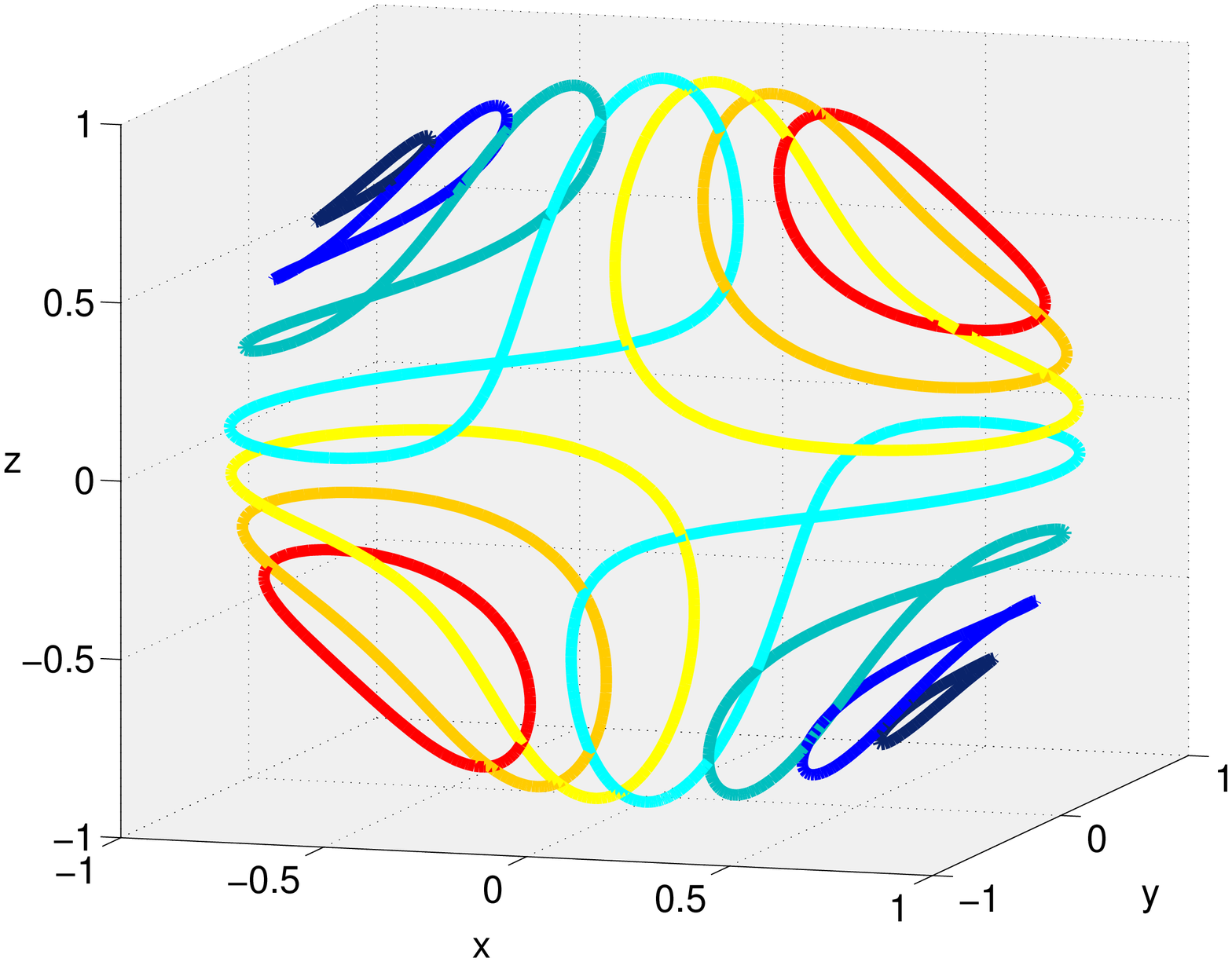}%
\end{minipage}

\caption{Same as Fig. \ref{fa}, but now the initial conditions for $\rho_{1}$
are given by $\gamma=\pi/4$ and $\varphi=\pi$. }

\label{fb} 
\end{figure}

We now calculate the trace distance between asymptotic reduced density
matrices corresponding to two different initial states of the QW without
decoherence while, in next Section, we extend the investigation to
the scenario that takes into account decoherence introduced by broken
links. Considering two different initial conditions given by Eqs.(\ref{psi0},\ref{phsi0}),
the difference between their asymptotic reduced density matrices is
\begin{equation}
\rho_{12}=\left(\begin{array}{cc}
\Pi_{1\mathit{L}}-\Pi_{2\mathit{L}} & \mathit{Q}_{10}-\mathit{Q}_{20}\\
\mathit{Q}_{10}^{\ast}-\mathit{Q}_{20}^{\ast} & \Pi_{1\mathit{R}}-\Pi_{2\mathit{R}}
\end{array}\right).
\end{equation}
Therefore, the distance between the asymptotic reduced density matrices
is defined as 
\begin{equation}
D(\rho_{12})=\frac{1}{2}tr\left\vert \rho_{12}\right\vert .
\end{equation}
After some algebra, taking into account Eq.(\ref{estacio}) with $\theta=\pi/4$,
the asymptotic trace distance can be expressed, in terms of the initial
conditions, as 
\begin{equation}
D(\rho_{12})=\sqrt{2\left[\mathfrak{Re}\left(\mathit{Q}_{10}-\mathit{Q}_{20}\right)\right]^{2}+\left[\mathfrak{Im}\left(\mathit{Q}_{10}-\mathit{Q}_{20}\right)\right]^{2}},\label{dist}
\end{equation}
where $\mathfrak{Re}\left(\mathit{Q}\right)$ ($\mathfrak{Im}\left(\mathit{Q}\right)$)
is the real (imaginary) part of $\mathit{Q}$, and $\mathit{Q}_{0}$
is given by Eq.(\ref{q0}). In order to study the dependence on the
initial conditions, we consider the evolution of pairs of independent
states under the QW map. 
We fix the initial conditions for the first state and study Eq.(\ref{dist})
by considering different points on the Bloch sphere as the initial
conditions for the second state. Figures \ref{fa} and \ref{fb} show
our results in two non-equivalent scenarios. As can be seen from these
figures, the asymptotic trace distance shows a non trivial behavior
as a function of the second state, once the first one is fixed. The
left panel can be used to get an idea on how much the trace distance
will be reduced (the minimum reduction being of the order of 1/2 in
the case represented in Fig. \ref{fa}, whereas lower values are reached
for the parameters that correspond to Fig. \ref{fb}). In fact, the
maximum value of $D(\rho_{12})$ can be shown to be reached when $\rho_{1}$
is defined by $\gamma=0$ and $\rho_{2}$ by $\gamma=\pi$ (or the
other way around): That is, when the two states are the North and
South poles of the Bloch sphere. As we will discuss later, the same
remains true when decoherence is introduced. 

The contour levels can be mapped to the points of the Bloch sphere
associated to the second state (right panel), thus providing a closer
relationship to physical states. As we see by comparing the two figures,
changing the first state does not translate into a simple rotation
of the Bloch sphere representation, the reason being that the coin
operator does not commute with arbitrary rotations. 

\section{Asymptotic density matrix with decoherence}

Now we study the dynamics of the reduced density matrix ($\overset{\sim}{\rho_{c}}$)
for the QW under the effect of decoherence. For this, we exploit the
model of decoherence, known as \textit{broken links}, that was proposed
for the first time in Ref. \cite{broken} and analyzed in the frame
of previous Section in Ref. \cite{Mostafa}. This model induces decoherence
in both degrees of freedom, coin and position. Similar results can
be found for other decoherence models.

At each time step $t$, the state of the links in the line is defined.
Each link has a probability $p$ of breaking in a given time step.
Clearly, for $p=0$, the ideal decoherence-free QW is recovered. During
the movement stage, if the walker is in a site with both the links
on right and left broken (this happens with probability $p^{2}$),
the walker does not move. With probability $(1-p)^{2}$ both links
are not broken and, in this case, the evolution normally occurs. With
probability $p(1-p)$ only one link is broken and the walker is \textit{forced}
to move to the other direction. Reference \cite{Mostafa} obtains
the superoperator $\mathit{L}_{k}$ that determines the dynamical
evolution of the QW with broken links 
\begin{equation}
\mathit{L}_{k}=\left(\begin{array}{cccc}
1 & 0 & 0 & 0\\
0 & 0 & e & f+p^{2}\\
0 & 0 & p^{2}-f & e\\
0 & 1-2p & -2g & -2h
\end{array}\right),\label{lkb}
\end{equation}
where 
\begin{eqnarray}
e & = & \left(1-p\right)^{2}\sin2k,\nonumber \\
f & = & \left(1-p\right)^{2}\cos2k,\nonumber \\
g & = & p\left(1-p\right)\sin k,\nonumber \\
h & = & p\left(1-p\right)\cos k.
\end{eqnarray}
The dynamics of the reduced density matrix is again determined by
Eq.(\ref{rr}) but now $\mathit{L}_{k}$ is given by Eq.(\ref{lkb}).
Redefining $\mathit{M}_{k}$ as 
\begin{equation}
\mathit{M}_{k}=\left(\begin{array}{ccc}
0 & e & f+p^{2}\\
0 & p^{2}-f & e\\
1-2p & -2g & -2h
\end{array}\right),
\end{equation}
it is easy to prove that its eigenvalues $\{\lambda_{i}/i=1,2,3\}$
satisfy $\left\vert \lambda_{i}\right\vert <1$ for $p\neq0$. If
$A$ is the matrix constructed from the eigenvectors of the matrix
$\mathit{M}_{k}$, and $\Lambda$ the diagonal matrix with the eigenvalues
as elements, it is straightforward to prove that 
\begin{eqnarray}
\mathit{\lim_{t\rightarrow\infty}M_{k}^{t}} & = & \lim_{t\rightarrow\infty}(A\Lambda^{t}A^{\dagger})=0.
\end{eqnarray}
In this case Eq.(\ref{rr}) gives us 
\begin{eqnarray}
\overset{\sim}{\rho}_{c} & = & \int\limits _{-\pi}^{\pi}\frac{dk}{2\pi}\left(\begin{array}{cccc}
1 & 0 & 0 & 0\\
0 & 0 & 0 & 0\\
0 & 0 & 0 & 0\\
0 & 0 & 0 & 0
\end{array}\right)\left(\begin{array}{c}
r_{0}\\
r_{1}\\
r_{2}\\
r_{3}
\end{array}\right).
\end{eqnarray}
In other words, in the formalism of $2\times2$ matrices, the reduced
density matrix in the asymptotic regime is simply 
\begin{equation}
\overset{\sim}{\rho_{c}}=\frac{1}{2}\left(\begin{array}{cc}
1 & 0\\
0 & 1
\end{array}\right),\label{opera2}
\end{equation}
regardless of the initial state. Thus, in the presence of noise, the
trace distance of any two different initial states approaches zero,
i.e. 
\[
lim_{t\rightarrow\infty}D(\rho_{12})=0.
\]

\section{Short-time behavior}

So far we have investigated the properties of the reduced density
matrix in the long-time regime. We obtained a definite limit for both
the decoherence-free scenario and the case with decoherence. We now
discuss the situation where one considers not the asymptotic limit
but a finite number of steps in the QW. Our study, as before, is focused
on the time evolution of $D(\rho_{1},\rho_{2})$.

The measure of non-Markovianity given by Ref. \cite{Breuer2009} is
based on the rate of change %
\footnote{For the QW considered here, $t$ takes only discrete values $t\in\mathbb{N}$,
therefore time derivatives and integrals in time have to be understood
as finite differences and sums.%
} of the trace distance 
\begin{equation}
\sigma(t,\rho_{1,2}(0))=\frac{d}{dt}D(\rho_{1}(t),\rho_{2}(t)).\label{derivdist}
\end{equation}
\begin{figure}
\includegraphics[width=8cm]{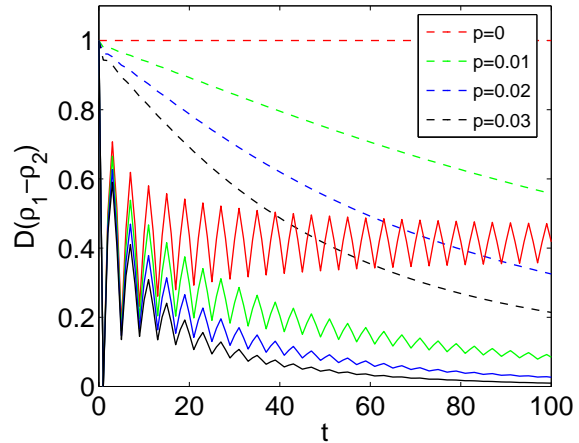} \caption{Trace distance, as a function of the number of time steps, between
the whole density matrices (dashed lines) and between the corresponding
reduced density matrices (solid lines). The initial state $\rho_{1}(0)$
is defined by Eq.(\ref{opera}) with $\gamma=0$, while $\rho_{2}(0)$
with $\gamma=\pi$. Different values of the decoherence parameter
$p$ have been considered.}

\label{figrho12_t} 
\end{figure}

Figure \ref{figrho12_t} shows the time evolution of the trace distance
both for the whole density matrices and for the corresponding reduced
density matrices associated with the pair of states giving the maximum
trace distance (see below). If one starts from a different pair of
states, the curves look qualitatively similar, although the overall
scale is smaller.

We have considered various values of the decoherence parameter $p$,
the case $p=0$ corresponding to the absence of decoherence. Without
decoherence, the QW evolves unitarily, so that the trace distance
between two total states is preserved. If $p>0$, the evolution for
the total state is clearly Markovian, as indicated by a monotonous
decrease in the trace distance (this happens of course for any possible
pair of initial states). The reduced density matrices, however, show
a completely different behavior. Considering first the case $p=0$,
we observe the presence of oscillations, implying that the trace distance
increases during some time intervals, giving a positive value of $\sigma$
in Eq. (\ref{derivdist}). As discussed in Ref. \cite{Breuer2009},
this feature is a clear signature of a non-Markovian process. We notice
that the amplitude of these oscillations decreases with $t$. 
For values $p>0$, we also observe the presence of such oscillations.
In fact, the curves look similar during the first time steps. However,
as $t$ increases the oscillations are more strongly damped than in
the decoherence-free case. This effect is even more pronounced for
larger values of $p$. In addition to these features, we also notice
that the trace distance goes asymptotically to zero, consistently
with our results in Sect. IV for the asymptotic limit.

\begin{figure}
\includegraphics[width=8cm]{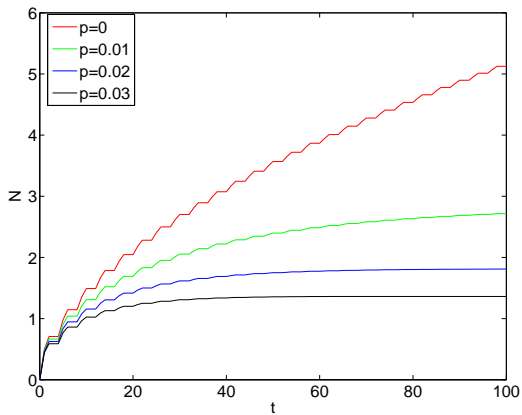}

\caption{Contribution to the non-Markovianity measure, as a function of the
number of time steps, evaluated for the pair of initial states $\rho_{1}(0)$
and $\rho_{2}(0)$ that maximizes the integral in Eq. (\ref{totalnonMark}).
Different values of the decoherence parameter $p$ have been considered.}

\label{AcNonMar1} 
\end{figure}

\begin{figure}
\includegraphics[width=8cm]{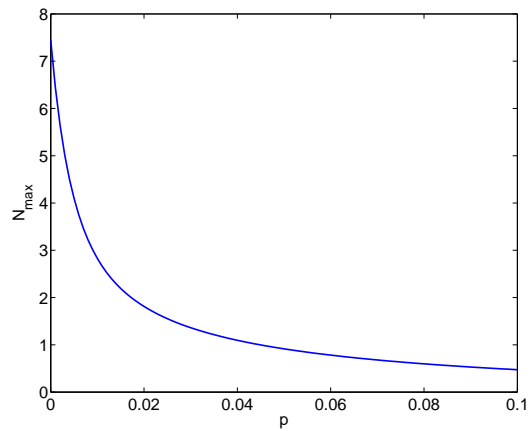}

\caption{The non-Markovianity measure, as defined by Eq. (\ref{totalnonMark}),
calculated for the time interval $[0,200]$ as a function of $p$,
for the pair of states that maximizes the final trace distance. }

\label{Nmaxp}
\end{figure}

To obtain a quantitative idea about the degree of the non-Markovianity
observed in the previous plots, the authors in Ref. \cite{Breuer2009}
suggest, as a figure of merit, the accumulated area of the trace distance
variation for those time intervals where the trace distance is increasing,
which amounts to calculating 
\begin{equation}
N_{max}=\max_{\rho_{1},\rho_{2}}\int_{\sigma>0}\sigma(\tau,\rho_{1,2}(0))d\tau.\label{totalnonMark}
\end{equation}
The maximization is performed over all the possible pairs of initial
states $\rho_{1}(0)$ and $\rho_{2}(0)$. In our case, due to the
fact that we are dealing with a two-level system (the coin), we can
restrict our investigation to pairs of orthogonal pure states. Even
though, a maximization procedure that takes into account a large time
interval is computationally hard. We have checked numerically that
the pair of states that maximizes Eq. (\ref{totalnonMark}), at least
within the time interval $[0,50]$ is the same as in the free case,
i.e. the North and South poles of the Bloch sphere. Given its simplicity,
we assume that this result holds even for larger time intervals. We
have therefore plotted in Fig. \ref{AcNonMar1} the value 
\begin{equation}
N(t)=\int_{\sigma>0;\tau\in[0,t]}\sigma(\tau,\rho_{1,2}(0))d\tau,\label{contribution}
\end{equation}
 evaluated for this pair of initial states. $N(t)$ can be seen as
the contribution to the non-Markovianity measure in the time window
$[0,t]$. Even if in the time window allowed by our computational
power it is not possible to evaluate (if any) the asymptotic value
of $N(t)$ for $t\rightarrow\infty$ (i.e. the non-Markovianity measure
$N_{max}$) in the decoherence-free case, the results reported in
Fig. \ref{AcNonMar1} give already a very precise picture of how the
decoherence affects the degree of non-Markovianity of the coin evolution.
This non-Markovianity is stronger as the magnitude of decoherence
decreases, with the largest value of its measure corresponding to
the decoherence-free case. 
Fig. \ref{Nmaxp} shows clearly this feature, where we have plotted
the non-Markovianity parameter $N_{max}$ calculated for the time
interval $[0,200]$ as a function of $p$, for the North-South pair
of states. The curve plotted on this figure can be approximated by
the fitting formula $N_{max}(p)\simeq\frac{7.32}{1+150\, p}$. 

\section{Conclusions}

In this work we have analyzed both the short-time behavior and the
asymptotic limit of the reduced (or chiral) density matrix (i.e. when
the spatial degree of freedom is traced out) for the discrete time
QW on a one-dimensional lattice. We have found that this reduced system
shows clear features which can be associated to a non-Markovian evolution.
First, we considered the case where the QW proceeds without decoherence.
We observed that the chiral density matrix possesses a well-defined
asymptotic limit in time. This allows us to calculate the limiting
value of the trace distance for pairs of different initial states.

We have studied the effect of decoherence, modeled as the random presence
of broken links on the lattice. The case with decoherence possesses
a trivial asymptotic limit, since all states converge to the identity,
so that the trace distance between pairs of them always tends to zero.

The short-time behavior of the reduced system features quite interesting
results. One observes the presence of oscillations in the trace distance
for reduced matrices that correspond to two different initial states,
a phenomenon that clearly indicates a non-Markovian time evolution.
These oscillations appear even when the system does not suffer from
decoherence, and they are damped as the number of time steps increases,
thus allowing for a convergence of the trace distance, in accordance
with our previous observations. As the level of noise becomes larger,
the amplitude of the oscillations is also reduced, for a given number
of time steps. In addition, the trace distance approaches asymptotically
zero, as already predicted from our long-time analysis. The contribution
to the non-Markovianity measure reported in Eq. (\ref{totalnonMark}),
as a function of the number of time steps, then tends to a value that
decreases as the level of decoherence increases.

To conclude, we have found and characterized a non-Markovian behavior
for a relatively simple and yet non-trivial system as the coin in
a QW on a line. The results that we have presented for the particular
model of decoherence chosen here can also be found for other models,
as the one investigated in Ref. \cite{alejo2013a}. They provide a
step forward in our understanding of phenomena like the transition
from unitary to diffusive processes and of the thermalization of quantum
systems, and clearly deserve further attention. 
\begin{acknowledgments}
This work has been supported by the Spanish Ministerio de Educación
e Innovación, MICIN-FEDER project FPA2011-23897 and ``Generalitat
Valenciana'' grant PROMETEO/2009/128. C.D.F. acknowledges the support
from the UK EPSRC, Grant No. EP/G004579/1 under the ``New directions
for EPSRC research leaders\textquotedbl{} initiative. A.R. acknowledges
the support from PEDECIBA, ANII. C.D.F. is thankful to A. Pérez and
the Universitat de València for the kind hospitality.\end{acknowledgments}


\begin{thebibliography}{10}
\bibitem{BreuerPetruccione} H. P. Breuer and F. Petruccione, \textit{The
Theory of Open Quantum Systems} (Oxford University Press, Oxford,
2002).

\bibitem{Aharonov} Y. Aharonov, L. Davidovich, and N. Zagury, \emph{Phys.
Rev. A} \textbf{48}, 1687 (1993).

\bibitem{alejo2003} A. Romanelli \textit{et al.}, \emph{Physics Letters
A}, \textbf{313}, 325, (2003).

\bibitem{alejo2004} A. Romanelli \textit{et al.}, \emph{Physica A},
\textbf{338}, 395 (2004).

\bibitem{Breuer2009} H. P. Breuer, E. M. Laine, and J. Piilo, \emph{Phys.
Rev. Lett.} \textbf{103}, 210401 (2009).

\bibitem{BreuerUses} T. J. G. Apollaro \textit{et al.}, \emph{Phys.
Rev. A} \textbf{83}, 032103 (2011); P. Rebentrost and A. Aspuru-Guzik,
\emph{J. Chem, Phys.} \textbf{134}, 101103 (2011); P. Haikka \textit{et
al.}, \emph{Phys. Rev. A} \textbf{84}, 031602(R) (2011); S. Lorenzo,
F. Plastina, and M. Paternostro, \emph{Phys. Rev. A} \textbf{87},
022317 (2013); J.-S. Tang \textit{et al.}, \emph{Europhys. Lett.}
\textbf{97}, 10002 (2012); B.-H. Liu \textit{et al.}, \emph{Nat. Phys.}
\textbf{7}, 931 (2011); B.-H. Liu \textit{et al.}, \emph{Sci. Rep.}
\textbf{3}, 1781 (2012); T. J. G. Apollaro \textit{et al.}, arXiv:1311.2045
(2013).

\bibitem{alejo2010} A. Romanelli, \emph{Phys. Rev. A} \textbf{81},
062349 (2010).

\bibitem{alejo2012} A. Romanelli, \emph{Phys. Rev. A} \textbf{85},
012319 (2012).

\bibitem{alejo2013a} A. Pérez and A. Romanelli, \emph{Journal of
Computational and Theoretical Nanoscience} \textbf{10}, 1 (2013).

\bibitem{alejo2013b} A. Romanelli and G. Segundo, \emph{Physica A},
\textbf{393}, 646 (2014).

\bibitem{Brun} T. Brun, H. Carteret, and A. Ambainis, \emph{Phys.
Rev. A} \textbf{67}, 032304 (2003).

\bibitem{Mostafa} M. Annabestani, S. J. Akhtarshenas, M. R. Abolhassani,
\emph{Phys. Rev. A} \textbf{81}, 032321 (2010).

\bibitem{broken} A. Romanelli \textit{et al.}, \emph{Physica A} \textbf{347},
137 (2005).\end{thebibliography}
\end{document}